\documentclass[10pt,twocolumn,twoside] {IEEEtran}  
\usepackage[numbers,sort&compress]{natbib}
\usepackage{amsfonts,amsmath,amsthm,latexsym,amssymb,amscd,mathrsfs}
\usepackage{graphicx}
\usepackage{algorithm}
\usepackage{algorithmic}        

\newcommand{\OUTPUT}{\item[\algorithmicoutput]}
\newcommand{\algorithmicoutput}{\textbf{Output:}}
\usepackage{stfloats}
\usepackage{url}
\usepackage{bm}
\usepackage{dsfont}
\usepackage{subfigure}
\usepackage{stfloats}

\makeatletter

\newcommand{\Rmnum}[1]{\expandafter\@slowromancap\romannumeral #1@}
\makeatother

\newtheorem{thm}{Theorem}
\newtheorem{lem}{Lemma}
\newtheorem{pro}{Proposition}

\newtheorem{defx}{Definition}

\newtheorem{remark}{Remark}
\begin{document}
%
\title{A New Analysis for Support Recovery with Block Orthogonal Matching Pursuit}
%
%
%

\author{{Haifeng Li}, {Jinming Wen}
\thanks{This work was partially supported by the National Natural Science
Foundation of China (grant nos. 11701157, 11601134, 11671122 and 11871248).
}

\thanks{H. Li is with the School of Mathematics and Information Sciences, Henan Normal University, Jianshe Road 46, Xinxiang, China (E-mail: lihaifengxx@126.com)}
\thanks{Jinming Wen is with the College of Information Science and Technology, and College of Cyber Security,
Jinan University, Guangzhou, 510632, China (E-mail:jinming.wen@mail.mcgill.ca).}
}

\markboth{This manuscript was submitted to IEEE Signal Processing Letters}
{}
%



\maketitle

\begin{abstract}
Compressed Sensing (CS) is a signal processing technique which can accurately recover sparse
signals from linear measurements with far fewer number of measurements than those required
by the classical Shannon-Nyquist theorem.
Block sparse signals, i.e., the sparse signals whose nonzero coefficients occur in few blocks,
arise from many fields. Block orthogonal matching pursuit (BOMP) is a popular greedy algorithm
for recovering block sparse signals due to its high efficiency and effectiveness.
By fully using the block sparsity of block sparse signals, BOMP can achieve very good
recovery performance.
This paper proposes a sufficient condition to ensure that BOMP can exactly recover
the support of block $K$-sparse signals under the noisy case.
This condition is better than existing ones.
\end{abstract}


%
\IEEEpeerreviewmaketitle

\section{Introduction}
Compressed sensing (CS) \cite{compressed,Donoho,  WenTSP1, WenTSP,Jinmingwen15} has attracted much
attention in recent years. Suppose that we have linear model $\mathbf{{y}}=\mathbf{Ax}+\mathbf{e}$,
where $\mathbf{y}\in\mathbb{R}^m$ is a measurement vector, $\mathbf{A}\in \mathbb{R}^{m\times n}$
is a sensing matrix, $\mathbf{x}\in\mathbb{R}^n$ is a $K$-sparse signal
(i.e., $|\makebox{supp}(\mathbf{x})|\leq K$, where supp($\mathbf{x})=\{i:x_i\neq0\}$ is the support
of $\mathbf{x}$ and $|\makebox{supp}(\mathbf{x})|$ is the cardinality of supp($\mathbf{x}$))
and $\mathbf{e}$ represents the measurement noise.
Then under some conditions on $\mathbf{A}$, CS can accurately recover the support of $\mathbf{x}$
based on $\mathbf{y}$ and $\mathbf{A}$.

In many fields \cite{chen1,chen2}, such as DNA microarrays \cite{DNA}, multiple measurement vector
problem \cite{MMV} and direction of arrival estimation \cite{DOA}, the nonzero entries of $\mathbf{{x}}$
occur in blocks (or clusters).
Such kind of signals are referred to as block sparse signals and are denoted as $\mathbf{x}_B$
in this paper.

To mathematically define $\mathbf{x}_B$, analogous to \cite{Eldar2},
we view $\textbf{x}_B\in\mathbb{R}^{n}$ as a concatenation of blocks
$\textbf{x}_B[\ell]\in\mathbb{R}^{d}$:
\begin{equation}\label{blockx}
\mathbf{x}_B=[\mathbf{x}_B^T[1] \quad \mathbf{x}_B^T[2]\cdots \mathbf{x}_B^T[M]]^T,
\end{equation}
where $\mathbf{x}_B[\ell]$ with $\ell\in\Omega:=\{1, 2, \cdots,M\}$
denotes the $\ell$th block of $\mathbf{x}_B$.
Then,
\begin{defx}\label{def1}(\cite{Eldar2})
A vector $\mathbf{x}_B\in \mathbb{R}^n$ is called block $K$-sparse if $\mathbf{x}_B[\ell]$ is
nonzero for at most $K$ indices $\ell$.
\end{defx}

Denote
\begin{align}\label{eq.1-22}
T =supp^B(\mathbf{x}_B):=\{\ell| \mathbf{x}_B[\ell]\neq \mathbf{0}_{d\times1}\}.
\end{align}
Then, by Definition \ref{def1}, we have $|T|\leq K$ and $T\subseteq \Omega$.

Similar to $\mathbf{x}_B$, we also represent $\mathbf{A}$ as a concatenation of column-blocks
$\mathbf{A}[\ell]$ of size $m\times d$, $\ell\in\Omega$, i.e.,
\begin{equation}\label{blockA}
\mathbf{A}=[\mathbf{A}[1]\quad\mathbf{A}[2]\cdots\mathbf{A}[M]].
\end{equation}

Since block sparse signals arise from many fields \cite{chen1}, this paper focus on
study the recovery of $\mathbf{x}_B$ from measurements
\begin{align}\label{blockmodel}
\mathbf{y}=\mathbf{A}\mathbf{x}_B+\mathbf{e}.
\end{align}
To this end, we introduce the definition of block restricted isometry property (RIP).
\begin{defx}\label{def2}(\cite{Eldar2,BOGA1})
A matrix $\mathbf{A}$ has block RIP with parameter $\delta^B\in(0,1)$ if
\begin{equation}\label{blockRIP}
(1-\delta^B)\|\mathbf{h}_B\|_2^2\leq\|\mathbf{A}\mathbf{h}_B\|_2^2\leq(1+\delta^B)\|\mathbf{h}_B\|_2^2
\end{equation}
holds for every block $K$-sparse $\mathbf{h}_B\in\mathbb{R}^n$.
The minimum $\delta^B$ satisfying (\ref{blockRIP}) is defined as the block RIP constant $\delta^B_{K}$.
\end{defx}

To efficiently recover block sparse signals, the block OMP (BOMP) algorithm, which is described in Algorithm \ref{Alg}, has been proposed in \cite{Eldar2}.
Recently, using RIP,
\cite{Jinmingwen11} investigated some sufficient conditions for exact or stable recovery of block sparse signals with BOMP.
They also proved that their sufficient conditions are sharp in the noiseless case.
\begin{algorithm}
 \caption{The BOMP algorithm \cite{Eldar2}}
 \label{Alg}
 \begin{algorithmic}[1]
\REQUIRE $\mathbf{A}\in\mathbb{R}^{m\times n}$, $\mathbf{y}$,
\ENSURE $\mathbf{r}^0\leftarrow \mathbf{y}$, $k=1$, and $\Lambda^0=\emptyset$.
 \WHILE {``stopping criterion is not met"}
 \STATE Choose the block index $\lambda_k$ that satisfies\\
\qquad $\lambda_k=\arg \max\limits_{\ell \in \Omega}\| \mathbf{A}'{[\ell]}\mathbf{r}^{k-1} \|_2$. \label{alg.1}\\
 \STATE Let $\Lambda^k=\Lambda^{k-1}\bigcup \{\lambda_k\}$, 
 and calculate\\
$\qquad\mathbf{x}_B^k=\arg \min\limits_{\mathbf{x}_B: {supp}^B(\mathbf{x}_B)\subseteq \Lambda^k}
\| \mathbf{y}-\mathbf{A}\mathbf{x}_B\|_2$.\label{alg.3}\\
 \STATE 
 $\mathbf{r}^k=\mathbf{y}-\mathbf{y}^k=\mathbf{y}-\mathbf{A}\mathbf{x}_B^k$.\label{alg.2}\\
 \STATE $k\leftarrow k+1$.
 \ENDWHILE
  \OUTPUT $\mathbf{x}_B^{k}$ and $\Lambda^{k}$.
 \end{algorithmic}
 \end{algorithm}



In order to analyze the recoverability of BOMP in the noisy case,
we investigate the sufficient condition of the support recovery of block sparse signals
with $K$ iterations of BOMP in the noisy case.
The condition reduces to that for the noiseless case when $\mathbf{e}=\mathbf{0}$
and it is the results presented in \cite{Jinmingwen11}.

The rest of the paper is organized as follows.
We present our new sufficient conditions in Sections II and prove them in Sections III.
The paper is summarized in Section~IV.

\section{Main Results}

Similar to \cite{Jinmingwen11}, we define  mixed $\ell_2/\ell_p$-norm  as
\begin{align}\label{Beq.1}
\|\mathbf{x}_B\|_{2,p}=\|\mathbf{w}\|_p, \, p=1,2,\infty,
\end{align}
where $\mathbf{w}\in\mathbb{R}^{M}$ with $w_\ell=\|\mathbf{x}_B[\ell]\|_2$ $(\ell\in\Omega)$.
Then our sufficient condition for the exact support recovery of block $K$-sparse signals
with BOMP is as follows:
\begin{thm}
Suppose that in (\ref{blockmodel}), $\|\mathbf{e}\|_2\leq\varepsilon$ and $\mathbf{{A}}$
satisfies the block RIP of order $K+1$ with
\begin{align}
{\delta}^B_{K+1}<\frac{1}{\sqrt{{K}{}+1}}\label{sufcondition}.
\end{align}
Then BOMP with the stopping criterion $\|\mathbf{r}^k\|_2\leq\varepsilon$ can exactly recover
$T$ (see \eqref{eq.1-22}) from (\ref{blockmodel}) in $K$ iterations provided that
\begin{align}\label{Beq.5}
\min\limits_{i\in T}\|\mathbf{x}_B[i]\|_2>\frac{\varepsilon}{\sqrt{1-\delta^B_{K+1}}}+\frac{\varepsilon\sqrt{1+\delta^B_{K+1}}}{1-\sqrt{K+1}\delta^B_{K+1}}.
\end{align}
\label{thm1}
\end{thm}

The proof of Theorem \ref{thm1} will be given in Section \ref{Sec1}.

\begin{remark}\label{Rem-2}
\cite[Corollary 1]{Jinmingwen11} shows that if $\mathbf{A}$ and $\mathbf{e}$ in (\ref{blockmodel}) respectively satisfy the block RIP with $\delta_{K+1}^B$ satisfying (\ref{sufcondition}) and
$\|\mathbf{e}\|_2\leq\varepsilon$, then BOMP with the stopping criterion
$\|\mathbf{r}^k\|_2\leq\varepsilon$ exactly recovers $T$ (see \eqref{eq.1-22})
in $K$ iterations provided that
\begin{align}
\label{Beq.5exist}
\min\limits_{i\in T}\|\mathbf{x}_B[i]\|_2>\frac{2\varepsilon}{1-\sqrt{K+1}\delta_{K+1}^B}.
\end{align}

In the following, we show that our condition (\ref{Beq.5}) in Theorem \ref{thm1} is less restrictive
than \eqref{Beq.5exist}. Equivalently, we need to show that
\begin{align}\label{Beq.17}
\frac{2\varepsilon}{1-\sqrt{K+1}\delta_{K+1}^B}>\frac{\varepsilon}{\sqrt{1-\delta^B_{K+1}}}+\frac{\varepsilon\sqrt{1+\delta^B_{K+1}}}{1-\sqrt{K+1}\delta^B_{K+1}}.
\end{align}

Equivalently, we need to show
\begin{align}\label{Beq.18}
\big(2-\sqrt{1+\delta_{K+1}^B}\big)\sqrt{1-\delta_{K+1}^B}>{1-\sqrt{K+1}\delta_{K+1}^B}.
\end{align}

Since $1-\delta_{K+1}^B>1-\sqrt{K+1}\delta_{K+1}^B$, it is clear that (\ref{Beq.17}) holds if
\begin{align}\notag
\big(2-\sqrt{1+\delta_{K+1}^B}\big)\sqrt{1-\delta_{K+1}^B}>{1-\delta_{K+1}^B},
\end{align}
which is equivalent to
\begin{align}\label{Beq.20}
2-\sqrt{1+\delta_{K+1}^B}>\sqrt{1-\delta_{K+1}^B}.
\end{align}
It is easy to see that (\ref{Beq.20}) holds. Thus our condition is less restrictive
than \cite{Jinmingwen11}.

To clearly show the improvement of Theorem \ref{Beq.1} over \cite[Corollary 1]{Jinmingwen11},
we display $Z_1(K)-Z_2(K)$ versus $\delta^B_{K+1}$ for several $K$ in Figure \ref{fig:1},
where $Z_1(K)$ and $Z_2(K)$ respectively denote the right-hand sides of (\ref{Beq.5})
and \eqref{Beq.5exist}. From Figure \ref{fig:1}, we can see that
the improvement of Theorem \ref{Beq.1} over \cite[Corollary 1]{Jinmingwen11} is significant.
\begin{figure}[hbtp]
\centering
\includegraphics[width=0.5\textwidth]{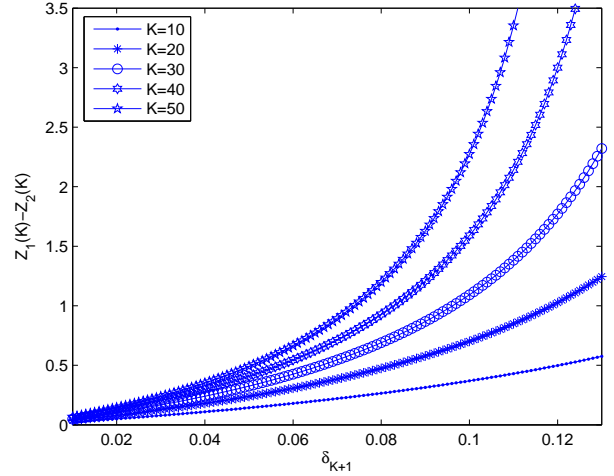}
\caption{The difference between $Z_1(K)$ and $Z_2(K)$.}
\label{fig:1}
\end{figure}
\end{remark}

\begin{remark}
We obtained a less restrictive sufficient condition for the exact support recovery of $K$-block sparse
signals with the BOMP algorithm based on RIC.
Since the weaker the RIC bound is, the less  number of measurements are needed.
The improved RIC results can be used in many CS-based applications, see, e.g., \cite{Song}.
\end{remark}

In the following,  we study the  worst-case necessity condition for the exact support recovery by BOMP. Recall that BOMP may fail to recover the support of $\mathbf{x}_B$ from $\mathbf{y}=\mathbf{A}\mathbf{x}_B$ if $\delta_{K+1}^B\geq\frac{1}{\sqrt{K+1}}$ \cite[Theorem 2]{Jinmingwen11}. Therefore,  $\delta_{K+1}^B<\frac{1}{\sqrt{K+1}}$ naturally becomes a necessity for the noisy case. Thus, we want to obtain the worst-case necessity condition on $\min\limits_{i\in T}\|\mathbf{x}_B[i]\|_2$ when $\delta_{K+1}^B<\frac{1}{\sqrt{K+1}}$.

\begin{thm}
Given any $\varepsilon>0$ and positive integer $K$. Let
\begin{align}\label{JBeq.50}
0<\delta<\frac{1}{\sqrt{K+1}}.
\end{align}
Then, there always exist a matrix $\mathbf{A}$ satisfying the RIP with
$\delta_{K+1}^B=\delta$, a block $K$-sparse vector $\mathbf{x}_B$ with
\begin{align}\label{JBeq.5}
\min\limits_{i\in T}\|\mathbf{x}_B[i]\|_2<\frac{\varepsilon}{\sqrt{1-(\delta_{K+1}^B)^2}(\sqrt{1-(\delta_{K+1}^B)^2}-\sqrt{K}\delta_{K+1}^B)},
\end{align}
and a noise vector  with $\|\mathbf{e}\|_2\leq\varepsilon$, such that BOMP
fails to recover $T$ (see \eqref{eq.1-22})  from (\ref{blockmodel}) in $K$ iterations.
\label{thm2}
\end{thm}
\proof
For any given positive integers $d$, $K$, and any real number $t_0>0$, $\varepsilon>0$, we construct a matrix function $\mathbf{A}(d)$, block $K$-sparse signal $\mathbf{x}_B$ and a noise vector $\mathbf{e}$. Let
\begin{equation}
\hspace{-2mm}\mathbf{{A}}(d)=\left(\begin{tabular}{llllll}
$\mathbf{I}_d$&$\mathbf{0}_{d\times (dK)}$ \\
$s\mathbf{E}_{(dK)\times d}$&$a\mathbf{I}_{dK}$\\
\end{tabular}
\right)_{d(K+1)\times d(K+1)},
\label{MEq.311}
\end{equation}

\begin{equation} \label{eq.36}
\mathbf{x}_B =
\left(\begin{tabular}{llllll}
$\mathbf{0}_{d \times 1}$ \\
$t_0\mathbf{e}_{1}$\\
$\vdots$\\
$t_0\mathbf{e}_{1}$\\
\end{tabular}
\right)_{d(K+1) \times 1}
\end{equation}
and
\begin{equation} \label{eq.361}
\mathbf{e} =
\left(\begin{tabular}{llllll}
$\varepsilon\mathbf{e}_{1}$ \\
$\mathbf{0}_{(dK) \times 1}$\\
\end{tabular}
\right)_{d(K+1) \times 1}
\end{equation}
where $\mathbf{I}_d$ being the $d\times d$ identity matrix, $\mathbf{0}_{d\times (dK)}\in\mathbb{R}^{d\times (dK)}$ with all of its entries being 0,
\begin{align}\label{atta1}
\mathbf{E}_{(dK)\times d}=(\mathbf{I}_d,\cdots,\mathbf{I}_d)'\in\mathbb{R}^{(dK)\times d},
\end{align}
\begin{align}\label{atta2}
s=\frac{\delta}{\sqrt{K}},\qquad a=\sqrt{1-\delta^2},
\end{align}
and $\mathbf{e}_1\in\mathbb{R}^d$ is the first
coordinate unit vector.
So, $\mathbf{x}_B$ is supported on $T= \{2,3,\cdots,K+1\}$, $T^c=\{1\}$ and $t_0=\min_{i\in T}\|\mathbf{x}_B[i]\|_2$.

By simple calculations we get
\begin{equation}
\mathbf{{A}}'(d)\mathbf{A}(d)=\left(\begin{tabular}{llllll}
$(1+Ks^2) \hspace{.5mm} \mathbf{I}_d$&$as\mathbf{E}'_{(dK)\times d}$ \\
$as\mathbf{E}_{(dK)\times d}$&$a^2\mathbf{I}_{dK}$\\
\end{tabular}
\right).
\label{eq.32}
\end{equation}

When $d=1$ and $K=1$, the eigenvalues $\{\lambda_i\}_{i=1}^2$ of $\mathbf{{A}}'(1)\mathbf{A}(1)$ are
\begin{align}\label{JBeq.21}
\lambda_1=1-\delta,\qquad\lambda_2=1+\delta.
\end{align}

When $d=1$ and $K>1$, the eigenvalues $\{\lambda_i\}_{i=1}^{K+1}$ of $\mathbf{{A}}'(1)\mathbf{A}(1)$ are
\begin{align}\label{JBeq.2}
&\lambda_i=1-\delta^2  \qquad 1\leq i\leq K-1,\notag\\
&\lambda_K=1+\delta, \qquad \lambda_{K+1}=1-\delta.
\end{align}

Thus, the RIP constant of $\mathbf{A}(1)$ is $\delta_{K+1}=\delta$ for $K\geq 1$.

In the following, we will show that the block RIP constant of $\mathbf{A}(d)$ is $\delta_{K+1}^B=\delta$.

Given any block $K$-sparse vector $\mathbf{w}_B\in\mathbb{R}^{d(K+1)}$. Let $\mathbf{u}$, $\mathbf{v}\in\mathbb{R}^{K+1}$ with $u_i=v_i=\|\mathbf{w}[i]\|_2$ for $2\leq i\leq K+1$ and $u_{1}=-v_{1}=\|\mathbf{w}_B[1]\|_2$. Then
\begin{align}
&\mathbf{w}_B'\mathbf{A}'(d)\mathbf{A}(d)\mathbf{w}_B=(1+Ks^2)\mathbf{w}_B'[1]\mathbf{w}_B[1]\notag\\
&\quad+2as\sum\limits_{i=2}^{K+1}\mathbf{w}_B'[i]\mathbf{w}_B[1]+a^2\sum\limits_{i=2}^{K+1}\mathbf{w}_B'[i]\mathbf{w}_B[i]\notag\\
&\leq(1+Ks^2)\|\mathbf{w}_B[1]\|_2^2+2as\sum\limits_{i=2}^{K+1}(\|\mathbf{w}_B[i]\|_2\|\mathbf{w}_B[1]\|_2)\notag\\
&\quad+a^2\sum\limits_{i=2}^{K+1}\|\mathbf{w}_B[i]\|_2^2\notag\\
&=(1+Ks^2)u_1^2+2as\sum\limits_{i=2}^{K+1}(u_iu_{1})+a^2\sum\limits_{i=2}^{K+1}u_i^2\notag\\
&=\mathbf{u}'\mathbf{A}'(1)\mathbf{A}(1)\mathbf{u}\leq(1+\delta)\|\mathbf{u}\|_2^2=(1+\delta)\|\mathbf{w}_B\|_2^2\label{JBEq.22}.
\end{align}

On the other hand, we have
\begin{align}
&\mathbf{w}_B'\mathbf{A}'(d)\mathbf{A}(d)\mathbf{w}_B=(1+Ks^2)\mathbf{w}_B'[1]\mathbf{w}_B[1]\notag\\
&\quad+2as\sum\limits_{i=2}^{K+1}\mathbf{w}_B'[i]\mathbf{w}_B[1]+a^2\sum\limits_{i=2}^{K+1}\mathbf{w}_B'[i]\mathbf{w}_B[i]\notag\\
&\geq(1+Ks^2)\|\mathbf{w}_B[1]\|_2^2-2as\sum\limits_{i=2}^{K+1}(\|\mathbf{w}_B[i]\|_2\|\mathbf{w}_B[1]\|_2)\notag\\
&\quad+a^2\sum\limits_{i=2}^{K+1}\|\mathbf{w}_B[i]\|_2^2\notag\\
&=(1+Ks^2)v_1^2+2as\sum\limits_{i=2}^{K+1}(v_iv_{1})+a^2\sum\limits_{i=2}^{K+1}v_i^2\notag\\
&=\mathbf{v}'\mathbf{A}'(1)\mathbf{A}(1)\mathbf{v}\geq(1-\delta)\|\mathbf{v}\|_2^2=(1-\delta)\|\mathbf{w}_B\|_2^2\label{JBEq.23}.
\end{align}

Combining (\ref{JBEq.22}) and (\ref{JBEq.23}), the block RIP constant of $\mathbf{A}(d)$ is
\begin{align}\label{JBEq.24}
\delta_{K+1}^B=\delta.
\end{align}

We now show that BOMP may fail to recover $T$ from
\begin{equation} \label{eq.3611}
\mathbf{y}=\mathbf{A}(d)\mathbf{x}_B+\mathbf{e}=
\left(\begin{tabular}{llllll}
$\varepsilon\mathbf{e}_{1}$ \\
$at_0\mathbf{e}_{1}$\\
$\vdots$\\
$at_0\mathbf{e}_{1}$\\
\end{tabular}
\right)_{d(K+1) \times 1}.
\end{equation}

Recall that the BOMP algorithm, in order to show this Theorem, we only need to show
\begin{align}\label{JBEq.27}
&a^2t_0=\|a^2t_0\mathbf{e}_1\|_2=\max_{i\in T}\|(\mathbf{A}(d)[i])'\mathbf{y}\|_2<\notag\\
&\max_{j\in T^c}\|(\mathbf{A}(d)[j])'\mathbf{y}\|_2=\|(\varepsilon+Kast_0)\mathbf{e}_1\|_2=\varepsilon+Kast_0.
\end{align}

By (\ref{JBeq.5}), it is easy to see that (\ref{JBEq.27}) holds.

This completes the proof.
\endproof

\begin{remark}
We may find the gap between the necessary condition and the sufficient condition is small. So, our sufficient condition is nearly optimal. In fact, for example, Let $K=10$ and $\delta_{K+1}^B=0.04$. The upper bound of (\ref{JBeq.5}) is $1.1468\varepsilon$, and the lower bound of (\ref{Beq.5}) is $2.1964\varepsilon$. The gap is $1.0496\varepsilon$.
\end{remark}

\section{Proof of Theorem \ref{thm1}}\label{Sec1}

By steps \ref{alg.3} and \ref{alg.2} of Algorithm \ref{Alg}, we have
\begin{align}
&\mathbf{r}^{k}=\mathbf{{y}}-\mathcal{P}_{\Lambda^{k}}\mathbf{{y}}=\mathcal{P}_{\Lambda^{k}}^\bot\mathbf{{y}}
\stackrel{(a)}{=}\mathcal{P}_{\Lambda^{k}}^\bot(\mathbf{{A}}[T]\mathbf{x}_B[T]+\mathbf{e})\notag\\
&=\mathcal{P}_{\Lambda^{k}}^\bot\mathbf{A}[T\setminus\Lambda^k]\mathbf{x}_B[T\setminus\Lambda^k]+\mathcal{P}_{\Lambda^{k}}^\bot\mathbf{e},\label{Beq.3}
\end{align}
where (a) follows from (\ref{blockmodel}) and supp$^B(\mathbf{x}_B)=T$. The symbol $\mathcal{P}_{\Lambda^{k}}=\mathbf{A}[\Lambda^k]
(\mathbf{A}'[\Lambda^k]\mathbf{A}[\Lambda^k])^{(-1)}\mathbf{A}'[\Lambda^k]$ denotes the orthogonal projection onto $\mathcal{R}(\mathbf{A}[\Lambda^{k}])$ that is the range space of $\mathbf{A}[\Lambda^{k}]$ and $\mathcal{P}_{\Lambda^{k}}^\perp=\mathbf{I}-\mathcal{P}_{\Lambda^{k}}$.

It is worth mentioning that the residual $\mathbf{r}^{k}$ is orthogonal to the columns of $\mathbf{{A}}[{\Lambda^k}]$, i.e.,
\begin{align}\label{Eqq.1061}
\|\mathbf{A}'[i]\mathbf{r}^k\|_2=\|\mathbf{A}'[i]\mathcal{P}_{\Lambda^k}^\bot\mathbf{{y}}\|_2=0,\qquad i\in\Lambda^{k}.
\end{align}

\subsection{Main Analysis}
The proof of Theorem \ref{thm1} is related to \cite{ChangLiu}. We will give a brief sketch for the proof of Theorem \ref{thm1}. Our proof consists of two steps. We show that BOMP chooses a correct index in each iteration in the first step. In the second step, we show that BOMP performs exactly $K$ iterations.

We prove the first step by induction. If BOMP selects a correct index at an iteration, we will say that BOMP makes a success at the iteration. First, we present the condition guaranteeing BOMP to make a success in the first iteration. Then, suppose that BOMP has been successful in the first $k$ iterations, we show that BOMP also makes its success in the $(k+1)$th iteration. Here, we assume $1\leq k<K$.

The proof for the first selection corresponds to the case
of $k=1$. Clearly the induction hypothesis $\Lambda^{k-1}\subseteq T$ holds
for this case since $\Lambda^{k-1}=\Lambda^0=\emptyset$.

If BOMP has been successful for the previous $k$ iterations, then it means that $\Lambda^k\subseteq T$ and $|T \cap \Lambda^k|=k$, In this sense, BOMP will make a success in the $(k+1)$th iteration, provided that $\lambda_{k+1}\in T$ (see Algorithm \ref{Alg}). Based on step \ref{alg.1} of Algorithm \ref{Alg} and (\ref{Eqq.1061}), in order to show that $\lambda_{k+1}\in T$, in the $(k+1)$th iteration, we need to show
\begin{align}\label{geq.7}
&\|\mathbf{A}'[T\setminus\Lambda^k]\mathbf{r}^k\|_{2,\infty}=\max_{i\in T\setminus\Lambda^k}\|\mathbf{A}'[i]\mathbf{r}^k\|_2\notag\\
&>\max_{j\in\Omega\setminus T}\|\mathbf{A}'[{j}]\mathbf{r}^k\|_2=\|\mathbf{A}'[\Omega\setminus T]\mathbf{r}^k\|_{2,\infty}.
\end{align}

From (\ref{geq.7}), for any $j\in\Omega\setminus T$, it suffices to show
\begin{align}\label{geq.22}
\|\mathbf{A}'[T\setminus\Lambda^k]\mathbf{r}^k\|_{2,\infty}-\|\mathbf{A}'[{j}]\mathbf{r}^k\|_2>0.
\end{align}

\subsection{Proof of inequality (\ref{geq.22})}
In this subsection, we will show that (\ref{geq.22}) holds for $1\leq k<K$ when (\ref{sufcondition}) and (\ref{Beq.5}) hold.

Suppose that
\begin{align}\label{geq.14}
\mathcal{P}_T\mathbf{{y}}=\mathbf{{A}}[T]\boldsymbol{\xi}_B[T]
\end{align}
with $\boldsymbol{\xi}_B\in\mathbb{R}^{Md}$ and supp$(\boldsymbol{\xi}_B)=T$.
For simplicity, we denote
\begin{align}\label{Beq.21}
\boldsymbol{\alpha}=\boldsymbol{\xi}_B[{T\setminus\Lambda^k}].
\end{align}

By (\ref{geq.14}), using the Cauchy-Schwarz inequality, we can have
\begin{align}
&\|\mathbf{A}'[T\setminus\Lambda^k]\mathbf{r}^k\|_{2,\infty}=\frac{\|\mathbf{A}'[T\setminus\Lambda^k]\mathbf{r}^k\|_{2,\infty}\|\boldsymbol{\alpha}\|_{2,1}}{\|\boldsymbol{\alpha}\|_{2,1}}\notag\\
&\stackrel{(a)}{\geq}\frac{\sum\limits_{i\in T\setminus\Lambda^k}\|\mathbf{A}'[i]\mathbf{r}^k\|_{2}\|\boldsymbol{\xi}_B{[i]}\|_{2}}{\|\boldsymbol{\alpha}\|_{2,1}}\stackrel{(b)}{\geq}\frac{\big\langle\mathbf{r}^k,\sum\limits_{i\in T\setminus\Lambda^k}\mathbf{A}[i]\boldsymbol{\xi}_B{[i]}\big\rangle}{\|\boldsymbol{\alpha}\|_{2,1}}\notag\\
&=\frac{\big\langle\mathbf{r}^k,\mathcal{P}_{\Lambda^k}^\bot(\mathbf{y}-\mathcal{P}_{T}^\bot\mathbf{y})\big\rangle}{\|\boldsymbol{\alpha}\|_{2,1}}
\stackrel{(c)}{=}\frac{\|\mathbf{r}^k\|_2^2-\big\langle\mathcal{P}_{\Lambda^k}^\bot\mathbf{{y}},\mathcal{P}_{\Lambda^k}^\bot\mathcal{P}_{T}^\bot\mathbf{y}\big\rangle}{\|\boldsymbol{\alpha}\|_{2,1}}\notag\\
&\stackrel{(d)}{=}\frac{\|\mathbf{r}^k\|_2^2-\big\langle\mathbf{{y}},\mathcal{P}_{T}^\bot\mathbf{y}\big\rangle}{\|\boldsymbol{\alpha}\|_{2,1}}\stackrel{}{=}\frac{\|\mathbf{r}^k\|_2^2-\|\mathcal{P}_{T}^\bot\mathbf{e}\|_2^2}{\|\boldsymbol{\alpha}\|_{2,1}},\label{Beq.8}
\end{align}
where (a) follows from (\ref{Beq.1}), (b) follows from Cauchy-Schwarz inequality, (c) is from $\mathbf{r}^k=\mathcal{P}_{\Lambda^k}^\bot\mathbf{y}$, (d) $\mathcal{P}_{\Lambda^k}^\bot\mathcal{P}_{T}^\bot=\mathcal{P}_{T}^\bot$.

Now, we can present a lower bound for left-hand-side of (\ref{geq.22}).
\begin{align}
&\|\mathbf{A}'[T\setminus\Lambda^k]\mathbf{r}^k\|_{2,\infty}-\|\mathbf{A}'[{j}]\mathbf{r}^k\|_2\notag\\
&\geq\frac{\|\mathbf{r}^k\|_2^2-
\|\mathcal{P}_{T}^\bot\mathbf{e}\|_2^2}{\|\boldsymbol{\alpha}\|_{2,1}}-\|\mathbf{A}'[{j}]\mathbf{r}^k\|_2=\eta\label{Beq.12}.
\end{align}
So, to show (\ref{geq.22}), we only need to show $\eta>0$.

\begin{pro}\label{pro1}
Define $\mathbf{h}\in\mathbb{R}^{d}$ with
  \begin{align}\label{geq.121}
 \mathbf{h}=\frac{\mathbf{A}'[j]\mathcal{P}_{\Lambda^k}^\bot\mathbf{y}}
 {\|\mathbf{A}'[j]\mathcal{P}_{\Lambda^k}^\bot\mathbf{y}\|_2},
  \end{align}
for $j\in \Omega\setminus T$. We have $\|\mathbf{h}\|_2=1$.
Define
\begin{align}\label{geq.91}
\mathbf{B}=\mathcal{P}_{\Lambda^k}^\bot[\mathbf{{A}}[{T\setminus\Lambda^k} ] ~\mathbf{{A}}[j]],
\end{align}
\begin{equation}\label{geq.101}
  \mathbf{u}=\left[
   \begin{aligned}
   &\boldsymbol{\alpha}\\
   &\mathbf{0} \\
   \end{aligned}
   \right]\in\mathbb{R}^{|T\setminus\Lambda^k|d+d},
   \mathbf{v}=\left[
   \begin{aligned}
   &\mathbf{0}\\
   &\mathbf{h} \\
   \end{aligned}
   \right]\in\mathbb{R}^{|T\setminus\Lambda^k|d+d},
  \end{equation}
  where $\boldsymbol{\alpha}$ is defined in (\ref{Beq.21}).
For any $t>0$, we have
\begin{align}\label{Beq.9}
\eta=&\frac{1}{4t}\|\mathbf{B}((t+\frac{1}{\|\boldsymbol{\alpha}\|_{2,1}})\mathbf{u}-\mathbf{v})\|_2^2\notag\\
&-\frac{1}{4t}\|\mathbf{B}((t-\frac{1}{\|\boldsymbol{\alpha}\|_{2,1}})\mathbf{u}+\mathbf{v})\|_2^2-\mathbf{e}'\mathcal{P}_{T}^\bot\mathbf{A}[j]\mathbf{h},
\end{align}
where $\eta$ is defined in (\ref{Beq.12}).
\end{pro}
The proof of Proposition \ref{pro1} will be given in Section \ref{secappendixA}.

By the property of block RIP, it follows that
\begin{align}
&\|\mathbf{B}((t+\frac{1}{\|\boldsymbol{\alpha}\|_{2,1}})\mathbf{u}-\mathbf{v})\|_2^2-\|\mathbf{B}((t-\frac{1}{\|\boldsymbol{\alpha}\|_{2,1}})\mathbf{u}+\mathbf{v})\|_2^2\notag\\
&\stackrel{(a)}{\geq}(1-\delta_{K+1}^B)\|(t+\frac{1}{\|\boldsymbol{\alpha}\|_{2,1}})\mathbf{u}-\mathbf{v}\|_2^2\notag\\
&\quad-(1+\delta_{K+1}^B)\|(t-\frac{1}{\|\boldsymbol{\alpha}\|_{2,1}})\mathbf{u}+\mathbf{v}\|_2^2\notag\\
&\stackrel{(b)}{=}\frac{4t\|\boldsymbol{\alpha}\|_{2,2}^2}{\|\boldsymbol{\alpha}\|_{2,1}}-2t^2\delta_{K+1}^B\|\boldsymbol{\alpha}\|_{2,2}^2-2\frac{\|\boldsymbol{\alpha}\|_{2,2}^2\delta_{K+1}^B}{\|\boldsymbol{\alpha}\|_{2,1}^2}-2\delta_{K+1}^B\notag\\
&=4t\Big(\frac{\|\boldsymbol{\alpha}\|_{2,2}^2}{\|\boldsymbol{\alpha}\|_{2,1}}-\frac{\delta_{K+1}^B}{2}(t\|\boldsymbol{\alpha}\|_{2,2}^2+\frac{1}{t}\big(\frac{\|\boldsymbol{\alpha}\|_{2,2}^2}{\|\boldsymbol{\alpha}\|_{2,1}^2}+1)\big)\Big),\label{Beq.14}
\end{align}
where (a) follows from \cite[Lemma 3]{Jinmingwen11}, (b) follows from (\ref{geq.101}). 

Applying arithmetic-geometric mean inequality to (\ref{Beq.14}),
\begin{align}
&\|\mathbf{B}((t+\frac{1}{\|\boldsymbol{\alpha}\|_{2,1}})\mathbf{u}-\mathbf{v})\|_2^2-\|\mathbf{B}((t-\frac{1}{\|\boldsymbol{\alpha}\|_{2,1}})\mathbf{u}+\mathbf{v})\|_2^2\notag\\
&\geq\max\limits_{t>0}\Big\{4t\Big(\frac{\|\boldsymbol{\alpha}\|_{2,2}^2}{\|\boldsymbol{\alpha}\|_{2,1}}-\frac{\delta_{K+1}^B}{2}(t\|\boldsymbol{\alpha}\|_{2,2}^2+\frac{1}{t}\big(\frac{\|\boldsymbol{\alpha}\|_{2,2}^2}{\|\boldsymbol{\alpha}\|_{2,1}^2}+1)\big)\Big)\Big\}\notag\\
&=4t\|\boldsymbol{\alpha}\|_{2,2}\Big(\frac{\|\boldsymbol{\alpha}\|_{2,2}}{\|\boldsymbol{\alpha}\|_{2,1}}-{\delta_{K+1}^B}\sqrt{1+\frac{\|\boldsymbol{\alpha}\|_{2,2}^2}{\|\boldsymbol{\alpha}\|_{2,1}^2}}\Big)\label{Beq.15}
\end{align}

It follows from (\ref{Beq.12}) and (\ref{Beq.9}) that
\begin{align}
&\|\mathbf{A}'[T\setminus\Lambda^k]\mathbf{r}^k\|_{2,\infty}-\|\mathbf{A}'[{j}]\mathbf{r}^k\|_2\notag\\
&\geq\frac{1}{4t}(\|\mathbf{B}((t+\frac{1}{\|\boldsymbol{\alpha}\|_{2,1}})\mathbf{u}-\mathbf{v})\|_2^2-\|\mathbf{B}((t-\frac{1}{\|\boldsymbol{\alpha}\|_{2,1}})\mathbf{u}+\mathbf{v})\|_2^2)\notag\\
&\quad-\mathbf{e}'\mathcal{P}_{T}^\bot\mathbf{A}[j]\mathbf{h}\notag\\
&\stackrel{(a)}{\geq}\|\boldsymbol{\alpha}\|_{2,2}\Big(\frac{\|\boldsymbol{\alpha}\|_{2,2}}{\|\boldsymbol{\alpha}\|_{2,1}}-{\delta_{K+1}^B}\sqrt{1+\frac{\|\boldsymbol{\alpha}\|_{2,2}^2}{\|\boldsymbol{\alpha}\|_{2,1}^2}}-\frac{\sqrt{1+\delta_{K+1}^B}\|\mathbf{e}\|_2}{\|\boldsymbol{\alpha}\|_{2,2}}\Big)\notag\\
&\stackrel{(b)}{\geq}\|\boldsymbol{\alpha}\|_{2,2}\Big(\frac{1}{\sqrt{K-k}}-\delta_{K+1}^B\sqrt{1+(\frac{1}{\sqrt{K-k}})^2}\Big)\notag\\
&\quad-\|\boldsymbol{\alpha}\|_{2,2}\frac{\sqrt{1+\delta_{K+1}^B}\|\mathbf{e}\|_2}{\sqrt{K-k}\min\limits_{i\in T}\|\boldsymbol{\xi}_B[i]\|_2}\notag\\
&\stackrel{(c)}{>}\frac{\|\boldsymbol{\alpha}\|_{2,2}}{\sqrt{K-k}}\Big(\delta_{K+1}^B(\sqrt{K+1}-\sqrt{K-k+1})\Big)\geq 0,\label{Beq.13}
\end{align}
where (a) follows from (\ref{Beq.15}), $\|\mathbf{h}\|_2=1$ and \cite[Lemma 3]{Jinmingwen11}, (b) follows from the function $f(x)=x-\delta_{K+1}^B\sqrt{1+x^2}$ is monotonously increasing on the interval $[0,\infty)$, $\frac{\|\boldsymbol{\alpha}\|_{2,2}}{\|\boldsymbol{\alpha}\|_{2,1}}\geq\frac{1}{\sqrt{K-k}}$
and $\|\boldsymbol{\alpha}\|_{2,2}\geq\sqrt{K-k}\min\limits_{i\in T}\|\boldsymbol{\xi}_B\|_2$, (c) follows from Lemma \ref{lem8} (presented in Section \ref{secappendixB}) and (\ref{Beq.5}).


It remains to show that BOMP stops under the stopping rule $\|\mathbf{r}^k\|_2\leq\varepsilon$ when it performs exactly $K$ iterations. Hence, we need to prove $\|\mathbf{r}^k\|_2>\varepsilon$ for $0\leq k<K$ and $\|\mathbf{r}^K\|_2\leq\varepsilon$.

By (\ref{Beq.3}), for $0\leq k<K$, we have
\begin{align}
&\|\mathbf{r}^k\|_2=
\|\mathcal{P}_{\Lambda^{k}}^\bot\mathbf{A}[T\setminus\Lambda^k]\mathbf{x}_B[T\setminus\Lambda^k]+\mathcal{P}_{\Lambda^{k}}^\bot\mathbf{e}\|_2\notag\\
&\geq\sqrt{1-\delta_{K+1}^B}\|\mathbf{x}_B[T\setminus\Lambda^k]\|_2-\varepsilon\notag\\
&\stackrel{(a)}{\geq}\frac{\sqrt{1-\delta_{K+1}^B}\sqrt{1+\delta_{K+1}^B}\varepsilon}{1-\sqrt{K+1}\delta_{K+1}^B}\geq\frac{({1-\delta_{K+1}^B})\varepsilon}{1-\sqrt{K+1}\delta_{K+1}^B}\geq\varepsilon,\notag
\end{align}
where (a) follows from (\ref{Beq.5}).

Similarly, from (\ref{Beq.3}),
\begin{align}
&\|\mathbf{r}^K\|_2=\|\mathcal{P}_{\Lambda^{K}}^\bot\mathbf{A}[T\setminus\Lambda^K]\mathbf{x}_B[T\setminus\Lambda^K]+\mathcal{P}_{\Lambda^{K}}^\bot\mathbf{e}\|_2\notag\\
&\stackrel{(a)}{=}\|\mathcal{P}_{\Lambda^{K}}^\bot\mathbf{e}\|_2\leq\varepsilon,
\end{align}
where (a) is from $\Lambda^K=T$.
Thus, BOMP performs $K$ iteration.
\endproof

\section{Conclusion}
In this paper, in the noisy case, we have presented a sufficient condition,
which is weaker than existing ones, for the exact support recovery of block $K$-sparse signals
with $K$ iterations of BOMP.

%


\section{Proof of Proposition \ref{pro1}}\label{secappendixA}
\proof
Recall that (\ref{geq.14}) and (\ref{Beq.21}), we have
\begin{align}\label{geq.151}
&(t+\frac{1}{\|\boldsymbol{\alpha}\|_{2,1}})\mathcal{P}_{\Lambda^k}^\bot\mathbf{{y}}-\mathcal{P}_{\Lambda^k}^\bot\mathbf{{A}}[j]\mathbf{h}\notag\\
&=(t+\frac{1}{\|\boldsymbol{\alpha}\|_{2,1}})\mathcal{P}_{\Lambda^k}^\bot(\mathcal{P}_{T}\mathbf{{y}}+\mathcal{P}_{T}^\bot\mathbf{{y}})-\mathcal{P}_{\Lambda^k}^\bot\mathbf{{A}}[j]\mathbf{h}\notag\\
&=\mathbf{B}((t+\frac{1}{\|\boldsymbol{\alpha}\|_{2,1}})\mathbf{u}-\mathbf{v})+(t+\frac{1}{\|\boldsymbol{\alpha}\|_{2,1}})\mathcal{P}_{T}^\bot\mathbf{e}
\end{align}
and
\begin{align}\label{geq.161}
&(t-\frac{1}{\|\boldsymbol{\alpha}\|_{2,1}})\mathcal{P}_{\Lambda^k}^\bot\mathbf{{y}}+\mathcal{P}_{\Lambda^k}^\bot\mathbf{{A}}[j]\mathbf{h}\notag\\
&=(t-\frac{1}{\|\boldsymbol{\alpha}\|_{2,1}})\mathcal{P}_{\Lambda^k}^\bot(\mathcal{P}_{T}\mathbf{{y}}+\mathcal{P}_{T}^\bot\mathbf{{y}})+\mathcal{P}_{\Lambda^k}^\bot\mathbf{{A}}[j]\mathbf{h}\notag\\
&=\mathbf{B}((t-\frac{1}{\|\boldsymbol{\alpha}\|_{2,1}})\mathbf{u}+\mathbf{v})+(t-\frac{1}{\|\boldsymbol{\alpha}\|_{2,1}})\mathcal{P}_{T}^\bot\mathbf{e}
\end{align}

Using the property of norm and (\ref{geq.121}), we have
{\begin{align}
&\|(t+\frac{1}{\|\boldsymbol{\alpha}\|_{2,1}})\mathcal{P}_{\Lambda^k}^\bot\mathbf{{y}}-\mathcal{P}_{\Lambda^k}^\bot\mathbf{{A}}[j]\mathbf{h}\|_2^2\notag\\
&-\|(t-\frac{1}{\|\boldsymbol{\alpha}\|_{2,1}})\mathcal{P}_{\Lambda^k}^\bot\mathbf{{y}}+\mathcal{P}_{\Lambda^k}^\bot\mathbf{{A}}[j]\mathbf{h}\|_2^2\notag\\
&\stackrel{}{=}\frac{4t}{\|\boldsymbol{\alpha}\|_{2,1}}\|\mathbf{r}^k\|_2^2-4t\|\mathbf{A}'[j]\mathcal{P}_{\Lambda^k}^\bot\mathbf{{y}}\|_2\label{Beq.10}.
\end{align}}

\normalsize
On the other hand, according to
\begin{align}
&(t+\frac{1}{\|\boldsymbol{\alpha}\|_{2,1}})(\mathcal{P}_{T}^\bot\mathbf{e})'\mathbf{B}((t+\frac{1}{\|\boldsymbol{\alpha}\|_{2,1}})\mathbf{u}-\mathbf{v})\notag\\
&=(t+\frac{1}{\|\boldsymbol{\alpha}\|_{2,1}})(\mathbf{e})'(\mathcal{P}_{T}^\bot)'\mathbf{B}((t+\frac{1}{\|\boldsymbol{\alpha}\|_{2,1}})\mathbf{u}-\mathbf{v})\notag\\
&=-(t+\frac{1}{\|\boldsymbol{\alpha}\|_{2,1}})\mathbf{e}'\mathcal{P}_{T}^\bot\mathbf{A}[j]\mathbf{h}\label{MMME.1}
\end{align}
\normalsize
and
\begin{align}
&(t-\frac{1}{\|\boldsymbol{\alpha}\|_{2,1}})(\mathcal{P}_{T}^\bot\mathbf{e})'\mathbf{B}((t-\frac{1}{\|\boldsymbol{\alpha}\|_{2,1}})\mathbf{u}+\mathbf{v})\notag\\
&=(t-\frac{1}{\|\boldsymbol{\alpha}\|_{2,1}})\mathbf{e}'\mathcal{P}_{T}^\bot\mathbf{A}[j]\mathbf{h}\notag,
\end{align}
\normalsize
we obtain
\begin{align}
&\|\mathbf{B}((t+\frac{1}{\|\boldsymbol{\alpha}\|_{2,1}})\mathbf{u}-\mathbf{v})+(t+\frac{1}{\|\boldsymbol{\alpha}\|_{2,1}})\mathcal{P}_{T}^\bot\mathbf{e}\|_2^2\notag\\
&\quad-\|\mathbf{B}((t-\frac{1}{\|\boldsymbol{\alpha}\|_{2,1}})\mathbf{u}+\mathbf{v})+(t-\frac{1}{\|\boldsymbol{\alpha}\|_{2,1}})\mathcal{P}_{T}^\bot\mathbf{e}\|_2^2\notag\\
&=\|\mathbf{B}((t+\frac{1}{\|\boldsymbol{\alpha}\|_{2,1}})\mathbf{u}-\mathbf{v})\|_2^2-\|\mathbf{B}((t-\frac{1}{\|\boldsymbol{\alpha}\|_{2,1}})\mathbf{u}+\mathbf{v})\|_2^2\notag\\
&\quad+\frac{4t}{\|\boldsymbol{\alpha}\|_{2,1}}\|\mathcal{P}_{T}^\bot\mathbf{e}\|_2^2-4t\mathbf{e}'\mathcal{P}_{T}^\bot\mathbf{A}[j]\mathbf{h}\label{Beq.11}.
\end{align}

By (\ref{geq.151})-(\ref{Beq.10}) and (\ref{Beq.11}), it follows that
\begin{align}
&\|\mathbf{B}((t+\frac{1}{\|\boldsymbol{\alpha}\|_{2,1}})\mathbf{u}-\mathbf{v})\|_2^2-\|\mathbf{B}((t-\frac{1}{\|\boldsymbol{\alpha}\|_{2,1}})\mathbf{u}+\mathbf{v})\|_2^2\notag\\
&\quad+\frac{4t}{\|\boldsymbol{\alpha}\|_{2,1}}\|\mathcal{P}_{T}^\bot\mathbf{e}\|_2^2-4t\mathbf{e}'\mathcal{P}_{T}^\bot\mathbf{A}[j]\mathbf{h}\notag\\
&=\frac{4t}{\|\boldsymbol{\alpha}\|_{2,1}}\|\mathbf{r}^k\|_2^2-4t\|\mathbf{A}'[j]\mathcal{P}_{\Lambda^k}^\bot\mathbf{{y}}\|_2\notag.
\end{align}

After some manipulations, we can prove that (\ref{Beq.9}) holds.
\endproof
\section{Proof of Lemma 1}\label{secappendixB}
\begin{lem}\label{lem8}
Consider (\ref{blockmodel}) and (\ref{geq.14}). Suppose that $\|\mathbf{e}\|_2\leq\varepsilon$. Then we have
\begin{align}\notag
\min\limits_{i\in T}\|\boldsymbol{\xi}_B[i]\|_2\geq\min\limits_{i\in T}\|\mathbf{x}_B[i]\|_2-\frac{\varepsilon}{\sqrt{1-\delta^B_{K+1}}}.
\end{align}
\end{lem}
\proof
Define
\begin{align}
\mathcal{P}_T\mathbf{e}=\mathbf{A}[T]
(\mathbf{A}'[T]\mathbf{A}[T])^{(-1)}\mathbf{A}'[T]\mathbf{e}=\mathbf{A}[T]\boldsymbol{\theta}_B[T]\notag
\end{align}
with $\boldsymbol{\theta}_B\in\mathbb{R}^{Md}$ is a block $K$-sparse vector. Then, by using block RIP, we have
\begin{align}
\|\mathcal{P}_T\mathbf{e}\|_2=\|\mathbf{A}[T]\boldsymbol{\theta}_B[T]\|_2\geq\sqrt{1-\delta_{K+1}^B}\|\boldsymbol{\theta}_B\|_{2}.\notag
\end{align}

On the other hand, we can obtain $\|\mathcal{P}_T\mathbf{e}\|_2\leq \|\mathbf{e}\|_2\leq\varepsilon$. Then we can obtain
\begin{align}\label{Beq.7}
\|\boldsymbol{\theta}_B\|_2\leq\frac{\varepsilon}{\sqrt{1-\delta^B_{K+1}}}
\end{align}

From (\ref{geq.14}) and (\ref{blockmodel}), we have
\begin{align}
&\mathbf{{A}}[T]\boldsymbol{\xi}_B[T]=\mathcal{P}_T\mathbf{{y}}=\mathcal{P}_T(\mathbf{{A}}\mathbf{x}_B+\mathbf{e})\notag\\
&=\mathbf{A}[T]\mathbf{x}_B[T]+\mathcal{P}_T\mathbf{e}=\mathbf{A}[T](\mathbf{x}_B[T]+\boldsymbol{\theta}_B[T]).
\end{align}
So, we have $\boldsymbol{\xi}_B[T]=\mathbf{x}_B[T]+\boldsymbol{\theta}_B[T]$.

Thus, we can obtain
\begin{align}
&\min\limits_{i\in T}\|\boldsymbol{\xi}_B[i]\|_2=\min\limits_{i\in T}\|\mathbf{x}_B[i]+\boldsymbol{\theta}_B[i]\|_2\notag\\
&\geq\min\limits_{i\in T}\|\mathbf{x}_B[i]\|_2-\|\boldsymbol{\theta}_B[T]\|_{2}\notag\\
&\stackrel{(a)}{\geq}\min\limits_{i\in T}\|\mathbf{x}_B[i]\|_2-\frac{\varepsilon}{\sqrt{1-\delta^B_{K+1}}},
\end{align}
where (a) follows from (\ref{Beq.7}).
\endproof

\end{document}